\documentclass[a4paper,english]{scrartcl}
\usepackage{amsmath, amssymb, latexsym}
\usepackage[T1]{fontenc}
\usepackage[latin1]{inputenc}
\usepackage{bibentry}
\usepackage{lmodern} 
\usepackage{grffile}
\usepackage{pdfpages}
\usepackage[pdftex]{hyperref}
\usepackage{natbib}
\usepackage{lscape}
\usepackage{multicol}
\usepackage{ae}
\usepackage{color}
\usepackage{aecompl}
\usepackage[T1]{fontenc}
\usepackage[latin1]{inputenc}
\usepackage{amsmath}

\usepackage{graphicx}
  
\ifpdf
   \DeclareGraphicsExtensions{.pdf,.jpeg,.png}
\else
   \usepackage[dvips]{graphicx}
\fi

\usepackage{subfigure}
\usepackage{xcolor}
\usepackage{setspace}
\usepackage{appendix}
\usepackage{amssymb}
\usepackage{amsthm}
\usepackage{wrapfig}
\usepackage{bm}
\usepackage{fancybox} 
\usepackage{ifthen}

{\theoremstyle{plain}

}

\title{Differential Geometrically Consistent \\
Artificial Viscosity in Curvilinear Coordinates}
\author{Harald H\"oller\thanks{Department of Astro- and Particle Physics, University of Innsbruck, \href{mailto:harald.hoeller@uibk.ac.at}{harald.hoeller@uibk.ac.at}},  
Antti Koskela\thanks{Department of Mathematics, University of Innsbruck, \href{mailto:antti.koskela@uibk.ac.at}{antti.koskela@uibk.ac.at}}, 
Werner Benger\thanks{Center for Computation \& Technology at Louisiana State University, Baton Rouge, \href{mailto:werner@cct.lsu.edu}{werner@cct.lsu.edu}} 
Ernst Dorfi\thanks{Department of Astronomy, University of Vienna, \href{mailto:ernst.dorfi@univie.ac.at}{ernst.dorfi@univie.ac.at}}
}
\begin{document}
\maketitle 

\begin{abstract}
High-resolution numerical methods have been developed for nonlinear, discontinuous problems as they appear in simulations of astrophysical objects. One of the strategies applied is the concept of artificial viscosity. Grid-based numerical simulations ideally utilize problem-oriented grids in order to minimize the necessary number of cells at a given (desired) spatial resolution. We want to propose a modified tensor of artificial viscosity which is employable for generally comoving, curvilinear grids.We study a differential geometrically consistent artificial viscosity analytically and visualize a comparison of our result to previous implementations by applying it to a simple self-similar velocity field. We give a general introduction to artificial viscosity first and motivate its application in numerical analysis. Then we present how a tensor of artificial viscosity has to be designed when going beyond common static Eulerian or Lagrangian comoving rectangular grids. We find that in comoving, curvilinear coordinates the isotropic (pressure) part of the tensor of artificial viscosity has to be modified metrically in order for it to fulfill all its desired properties. 
\end{abstract}

\section{Introduction}\label{Intro}

In astrophysics a multitude of systems and configurations are described 
with concepts from hydrodynamics, often combined with gravitation, 
radiation and/or magnetism. Mathematically radiation 
hydrodynamics (RHD) and magnetohydrodynamics (MHD) are described by systems of 
coupled nonlinear partial differential equations. 
The Euler equations of hydrodynamics, the Maxwell equations as well as 
radiative transport equations are hyperbolic PDEs that connect certain
densities and fluxes via conservation laws. 
The numerical solutions of these equations essentially need 
to comprise this quality. Today there exists a wide range of numerical 
methods for conservation laws that 
ensure the conservation of mass, momentum, energy etc. if applied 
properly, and multiple fields in physics and astrophysics have adopted these 
sophisticated numerical methods for studying various applications. 

Standard numerical schemes for partial differential equations are 
established under the assumption of classical differentiability. 
Routine finite difference schemes of first order usually smear or 
smoothen the solution 
in the vicinity of discontinuities as they come with intrinsic numerical 
viscosity. Standard second order methods show something to the effect of 
the Gibbs phenomenon, where oscillations around shocks emerge.  
In the past decades so called high-resolution methods
have been developed 
in order to achieve proper accuracy and resolution for nonlinear, discontinuous 
problems as they appear also in RHD or MHD. 
One of these strategies is the concept of artificial viscosity 
which we will also briefly motivate in subsection \ref{Rankine_and_Entropy}. 

In higher-dimensional problems 
this artificial viscosity emerges as a tensorial quantity. We will 
demonstrate this in subsection \ref{Artificial_Viscosity_and_Flux}. 
The result we want 
to present in this paper can be seen as a tensor analytical 
consequence of the artificial viscosity in general curvilinear 
coordinates when using consistent metric tensors. 
In section \ref{Artificial_Viscosity} we will propose a correction 
for the commonly used tensor of artificial viscosity for curvilinear grids. 

This correction is motivated by astrophysical applications where one considers comoving nonlinear coordinates 
represented by non-conformal (non-angle preserving) maps from spherical coordinates. 
The authors are currently investigating the generation of grids that are asymptotically 
spherical but which allow certain asymmetries that can be found in 
rotating configurations, nonlinear pulsation processes etc. This new approach to grid-based 
astrophysical simulation techniques will be addressed extensively with numerical applications in a future paper. 

\begin{figure}
  \centering
  \subfigure[][]{\label{oblat:1}\includegraphics[width=0.35\hsize]{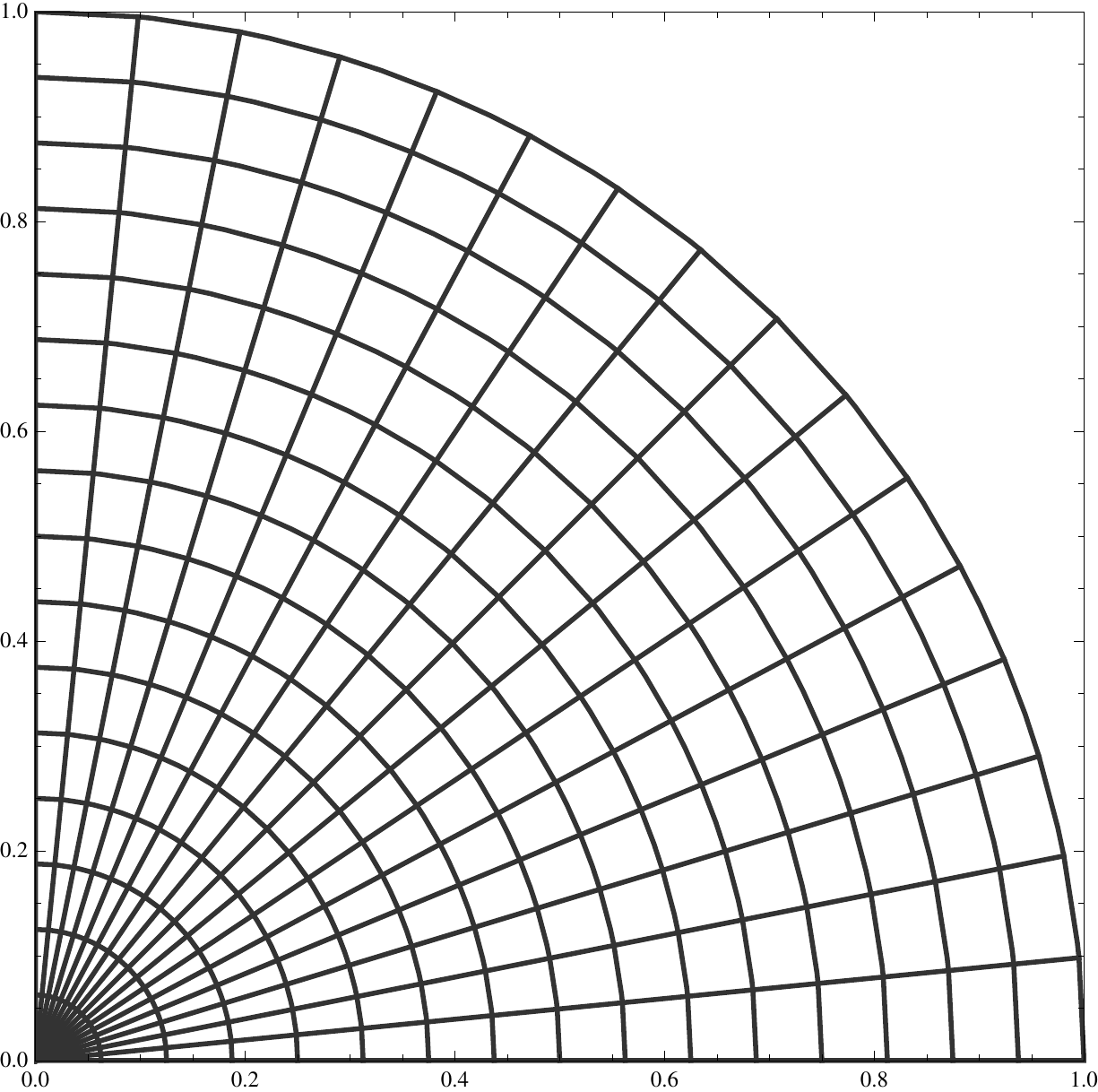}}   \qquad              
  \subfigure[][]{\label{oblat:2}\includegraphics[width=0.35\hsize]{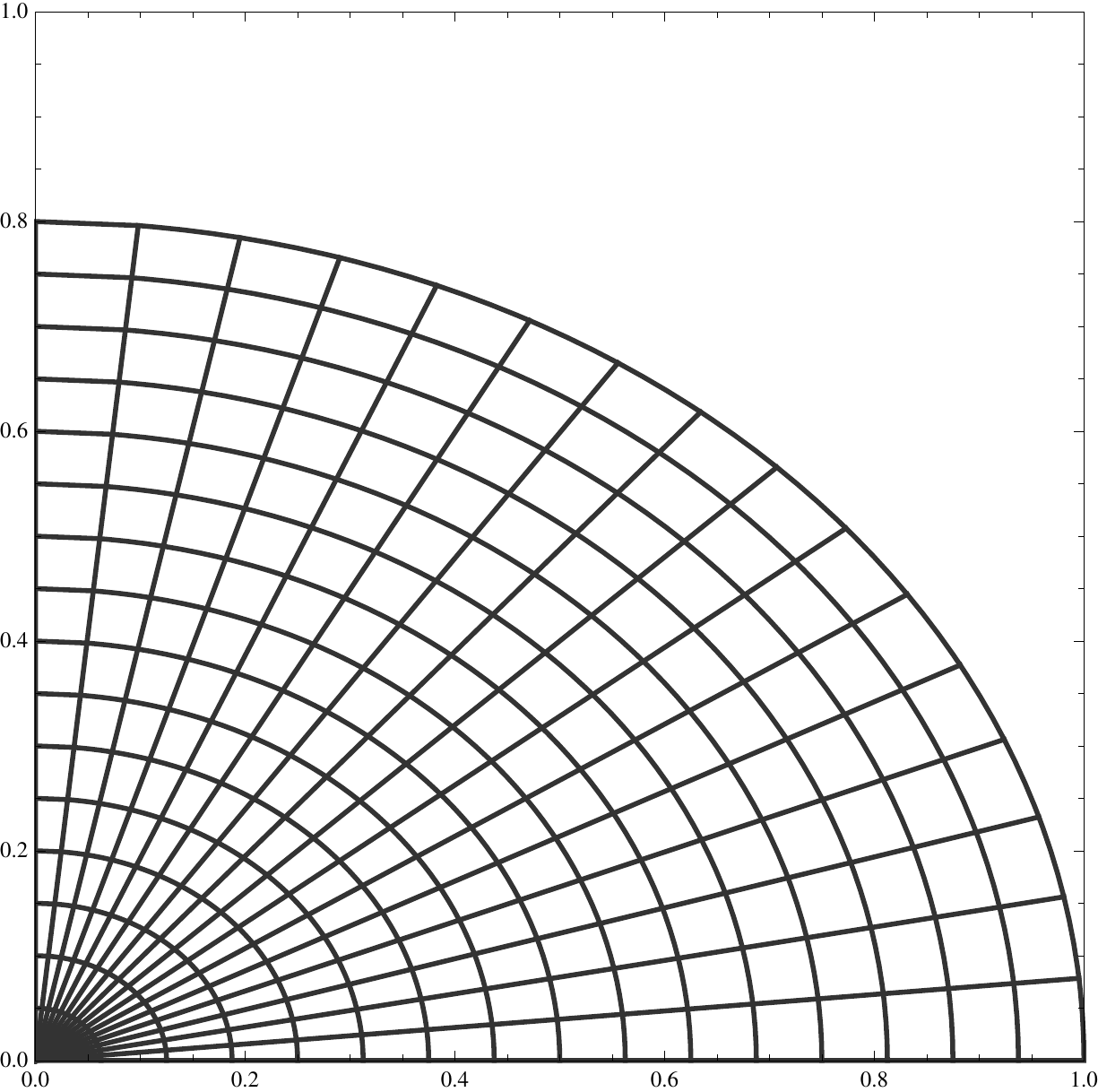}} \\
  \subfigure[][]{\label{oblat:3}\includegraphics[width=0.35\hsize]{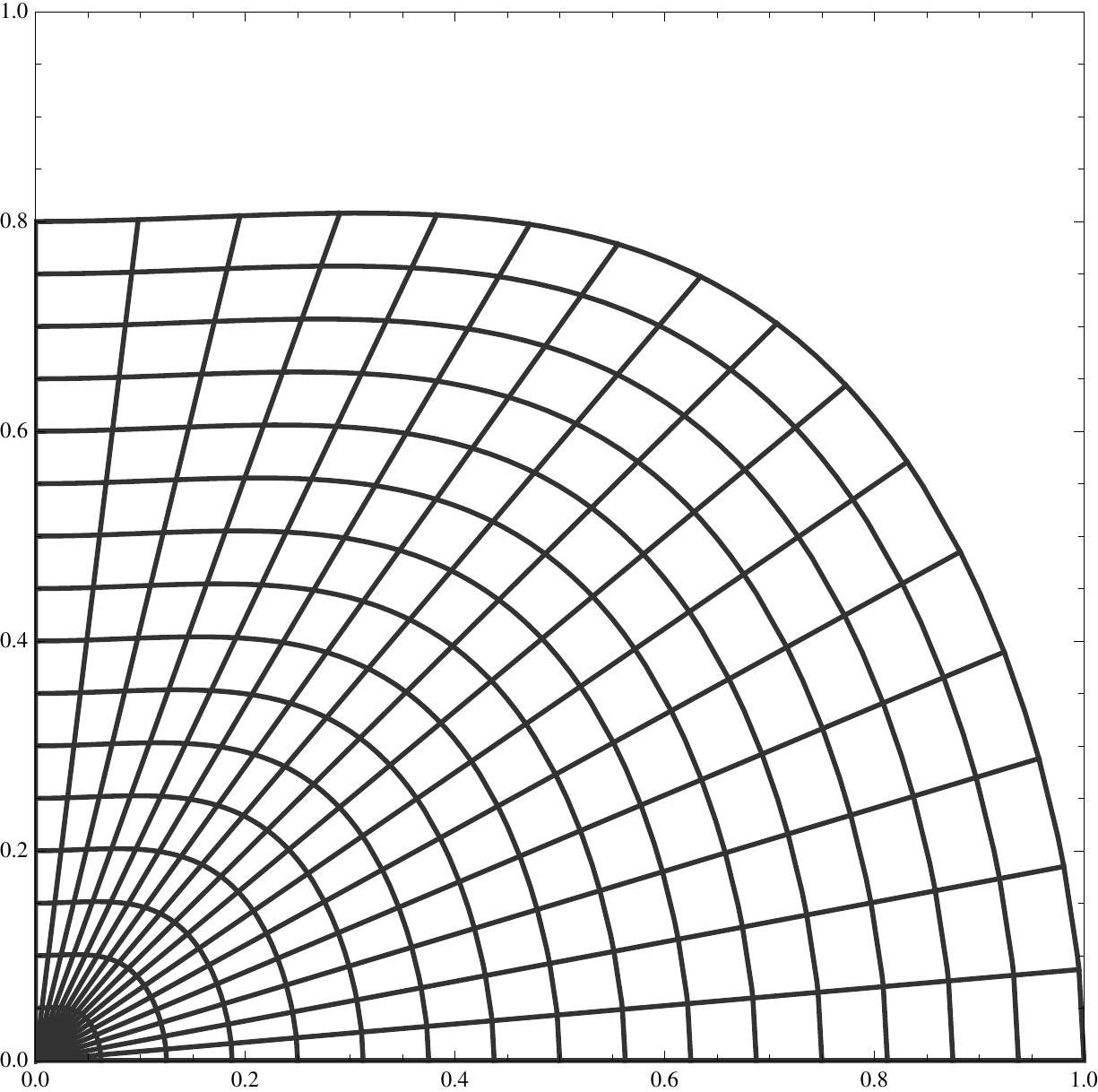}}   \qquad              
  \subfigure[][]{\label{oblat:4}\includegraphics[width=0.35\hsize]{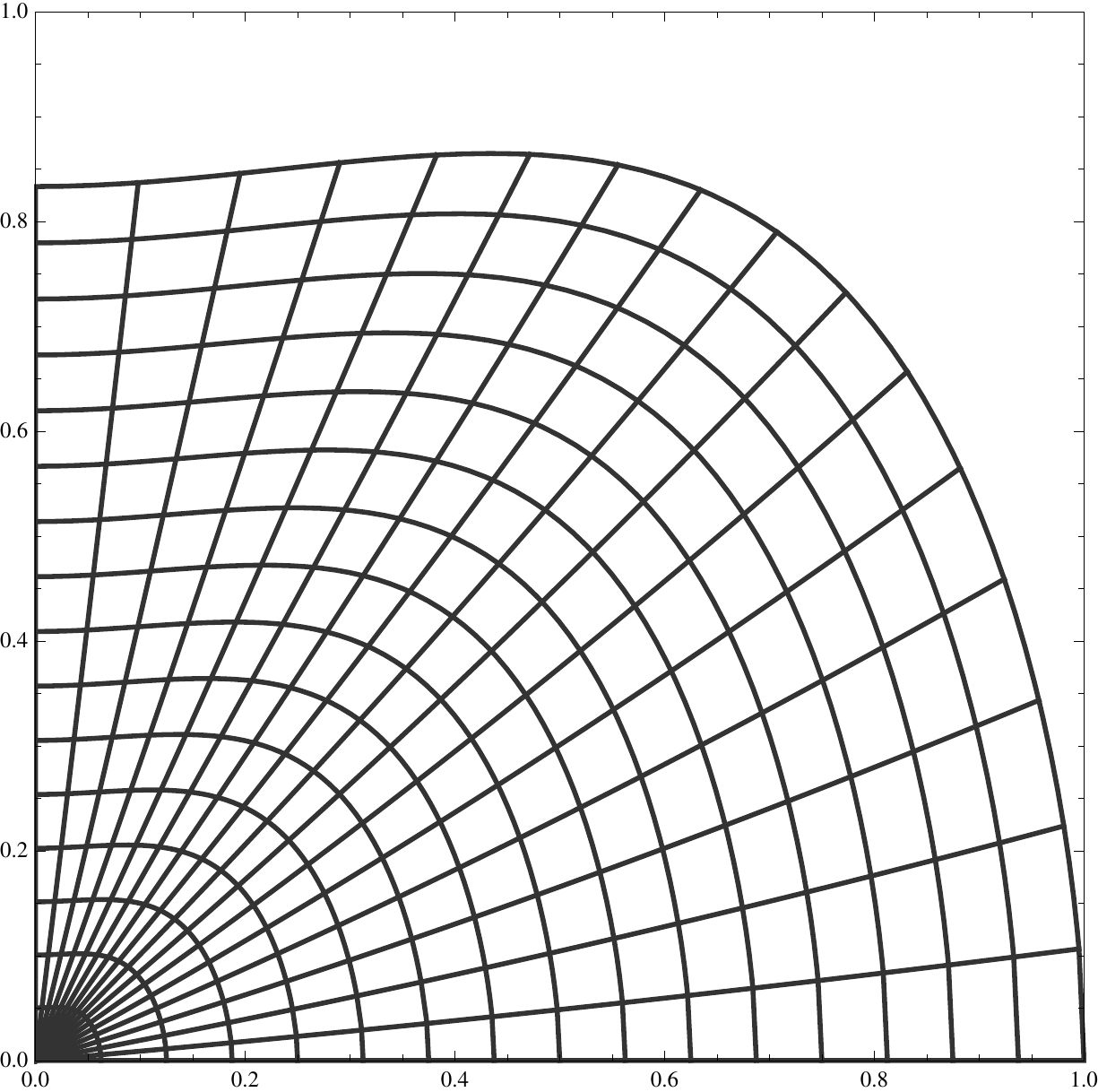}}
   \caption{Example of a non-conformal non-steady 2D grid with oblateness governed by three 
   choices of parameters $(a_1,a_2,a_3)$ as defined in equation (\ref{NonConformal_Grid}). }
  \label{Oblate_Grids}
\end{figure}

As an example of non-conformal two dimensional coordinates, Figure~\ref{Oblate_Grids} shows a grid that corresponds 
to the map $(x,y) \to (\xi, \eta)$,
\begin{equation}
  \label{NonConformal_Grid}
 \begin{aligned}
  &  x  =  \xi \cos \eta \\
  &  y =  (a_1 \xi + a_2 \xi ^2) \Big(1+ \frac{ a_3 \pi ^3-16 a_2 \xi + a_2 a_3 \pi ^3 \xi  }{4 \pi  (1+a_2 \xi )}\eta \,\, + \\
   &   + \, \frac{ 4 a_2 \xi -a_3 \pi^3 - a_2 a_3 \pi ^3 \xi }{\pi ^2 (1+b_2 \xi )}\eta ^2  +a_3 \eta^3 \Big) \sin \eta
 \end{aligned}
\end{equation} 
which gives the polar coordinates for the choice of parameters $(a_1,a_2,a_3) = (1,0,0).$ 
In such a nonorthogonal grid the metric tensor is no longer 
diagonal and one has to consider a consistent differential 
geometric approach to the formulation of the governing equations of RHD and MHD,
and also to the mathematical formulation of the artificial viscosity, which will be stressed in Section \ref{Artificial_Viscosity}.

The benefit of the consistent formulation can be seen when we consider time-dependent 
grids, e.g. when using time-dependent parameters $(a_1,a_2,a_3)$ in \eqref{NonConformal_Grid}. 
We refer to the Appendix \ref{appendix} for the depiction of the system 
of equations of RHD for generally comoving curvilinear coordinates with 
time-dependent metrics correspondingly.

\subsection{Brief Introduction to Conservation Laws} \label{Rankine_and_Entropy}

For the sake of stringency we recapitulate some important results from the 
theory and numerics of conservation laws which can be found e.g. in~\citep{leveque_1991} or
~\citep{richtmyer+morton_1994}. 

The equations of RHD and MHD
form a system of hyperbolic conservation laws that
describe the interaction of a density function $\mathbf{d}\left(\mathbf{x},t\right): 
\mathbb{R}^n \times \left[ 0,\infty \right) \rightarrow \mathbb{R}^m$ and its flux $\mathbf{f}\left( \mathbf{d} \right): \mathbb{R}^m \rightarrow \mathbb{R}^{m \times n}$. 
Equation (\ref{hd_cons_cs}) shows how a concrete choice for 
the density and the flux field can look like in a given coordinate system.  

The temporal change of the integrated density in a connected set $\Omega \subset \mathbb{R}^n$ then equals the flux over the boundary $\partial \Omega$, i.e.
\begin{equation} \label{erh} 
 \quad \partial_t \int\limits_{\Omega} \mathbf{d} \,\, \d V + \int\limits_{\partial \Omega} \mathbf{f} \cdot \mathbf{n} \, \d S = \mathbf{0} \quad \textrm{for all } \,\, t>0,
\end{equation}
where $\mathbf{n}$ is the outward oriented normal of the surface.

The system is called hyperbolic if the Jacobian matrix $\nabla_{\mathbf{d}}\mathbf{f}$ associated with the fluxes has real eigenvalues and if there exists a complete set of eigenvectors. In case of MHD and RHD this property has a direct physical relevance~\citep{pons+al_2000}. 

Assuming $\mathbf{f}$ to be a continuously differentiable function, equation \eqref{erh} can be rewritten via the divergence theorem as 
\begin{equation} \label{cauchy_problem_integral}
  \int\limits_t\int\limits_{\Omega} \big( \partial_t \mathbf{d} + \mathrm{div}_{\mathbf{x}} \, \mathbf{f}(\mathbf{d})\big) \,  \d V \, \d t = \mathbf{0}
\quad \textrm{ for all } \,\,  t > 0, \, \Omega \subset \mathbb{R}^n,
\end{equation}
which gives the system of partial differential equations for the density function $\mathbf{d}$:
\begin{equation} \label{pde_system}
 \partial_t \mathbf{d} + \mathrm{div}_{\mathbf{x}} \, \mathbf{f}(\mathbf{d}) = \mathbf{0} \quad \textrm{ for all } \,\,  t > 0, \, \mathbf{x} \in \mathbb{R}^n.
\end{equation} 
With an initial condition $\mathbf{d}(\mathbf{x},0)=\mathbf{d}_0(\mathbf{x})$, $\mathbf{x} \in \mathbb{R}^n$, this is called \it the Cauchy problem. \rm

In order to illustrate the connection of hydrodynamical applications to this formalism, we express the Euler equations in the 
form \eqref{pde_system}. The appearing variables are the gaseous density $\rho(\mathbf{x},t)$, the gas velocity $\mathbf{u}(\mathbf{x},t)$, 
the inner energy $\epsilon(\mathbf{x},t)$ and the gaseous pressure tensor $\mathbf{P}(\mathbf{x},t)$. Considering the 
differential form \eqref{pde_system}, we recognize the continuity equation, the equation of motion 
and the energy equation as the components of the hyperbolic problem. 
In case of the most relevant problem, that of 3D hydrodynamics, the density and its flux are given as
\begin{equation} \label{hd_cons}
 \mathbf{d} = \begin{pmatrix} \rho \vspace{1mm} \\ \rho\mathbf{u}\vspace{1mm}  \\ \rho \epsilon   \end{pmatrix}  \in \mathbb{R}^5, 
\quad \mathbf{f}(\mathbf{d}) = 
\begin{pmatrix} \rho \mathbf{u}^{\mathrm{T}} \vspace{1mm} \\ \rho \mathbf{u} \mathbf{u}^{\mathrm{T}} + \mathbf{P} \vspace{1mm} \\ \rho \epsilon \mathbf{u} + (\mathbf{P} \mathbf{u})^{\mathrm{T}}
\end{pmatrix} \in \mathbb{R}^{5 \times 3}.
\end{equation}

For a given coordinate system with base vectors $\mathbf{e}_i$, the tensorial fields are given explicitly as (using the Einstein notation)
\begin{equation} \label{hd_cons_cs}
 \mathbf{d} = \begin{pmatrix} \rho \vspace{1mm} \\ \rho u^i  \, \mathbf{e}_i \vspace{1mm}  \\ \rho \epsilon  \end{pmatrix} , 
\quad \mathbf{f}(\mathbf{d}) = \begin{pmatrix} \rho u^i \, \mathbf{e}_i^{\mathrm{T}}  \vspace{1mm} \\ \big( \rho u^i u^j  + P^{ij} \big) \, \mathbf{e}_i \mathbf{e}_j^{\mathrm{T}}
 \vspace{1mm} \\ \big( \rho \epsilon u^i + {P^i}_j u^j \big) \, \mathbf{e}_i^{\mathrm{T}}  \end{pmatrix}.
\end{equation}
The gaseous pressure tensor can be assumed to be isotropic in most applications, which means that $\mathbf{P} = g^{ij} p$ where $p(\mathbf{x},t)$ is 
the scalar gas pressure and $g^{ij}={\mathbf{e}^i}^{\mathrm{T}} \mathbf{e}^j$ the contravariant metric tensor. In case of adaptive grids respectively 
moving coordinate lines, the base vectors are time-dependent as well, i.e. then $\mathbf{e}_i = \mathbf{e}_i(\mathbf{x},t)$.

Since even the simplest examples of one-dimensional scalar conservation laws like the Burgers' equation have classical solutions only in some special 
cases, one has to broaden the considered function space of possible solutions. 
For the so-called weak solutions, we appeal to generalized functions where the discontinuities are defined properly. 
The generalized concept of differentiation of distributions shifts the operations to test functions $\gamma: \mathbb{R}^n \times \mathbb{R}^{+} \supset G 
\rightarrow \mathbb{R}$ ($G$ open) which are infinitely differentiable and have a compact support (meaning that for each $\gamma$ there exists a closed 
and bounded subset $K$ such that $\gamma (\mathbf{x},t)=0$  for all $x \, \in G \setminus K$). We denote this space of test functions by $D(G)$. 
In this generalized space of solutions the Cauchy problem (\ref{cauchy_problem_integral}) is written as
\begin{equation*}
 \int_{t \geq 0} \int_{\mathbb{R}^n} \big( \partial_t \mathbf{d} + \mathrm{div}_{\mathbf{x}} \, \mathbf{f}(\mathbf{u}) \big) \gamma \, \d V \, \d t = 0 \, 
\quad \textrm{for all } \, \gamma \in D(G).
\end{equation*}
The weak formulation of the conservation law (\ref{cauchy_problem_integral}) is obtained by shifting the derivatives to the test functions by partial integration,
and by using the compactness of the support. We get that the following has to hold for each $\gamma \in D(G)$:
\begin{equation} \label{cauchy_weak_formulation}
 \int\limits_{t \geq 0} \int\limits_{\mathbb{R}^n} \big( 
\mathbf{d} \, \partial_t \gamma + \mathbf{f}(\mathbf{d}) \, 
\nabla_{\mathbf{x}} \, \gamma  
\big) \, \d V \, \d t = - \int\limits_{\mathbb{R}^n} 
\gamma(\mathbf{x},0) \, \mathbf{d}_0(\mathbf{x}) \, \d V.
\end{equation}
The function $\mathbf{d} \in L^{\infty}$ is called a weak solution of the PDE \eqref{pde_system}, if it satisfies (\ref{cauchy_weak_formulation}) 
and $\mathbf{d} \in U$ with $\mathbf{d}_0 \in L^{\infty}$. However, there is a small drawback. This weak solution is not 
necessarily unique and usually further constraints have to be imposed in order to guarantee its uniqueness. This leads us 
to the actual topic of this paper.

\subsection{Introduction to Artificial Viscosity} \label{subsection:intro_art_vis}

For most physical problems it is naturally sufficient to look for weak solutions from the function space of piecewise continuously 
differentiable functions. Constraining the space of solutions in this way, we call the physical variables $\mathbf{d}$ weak 
solutions of the Cauchy problem \eqref{pde_system}, if they are classical solutions wherever they are continuously differentiable, and 
if at discontinuities (shocks) they satisfy additional conditions in order to be physically reasonable (we elaborate on these conditions
below).

The mathematical theory provides several techniques to distinguish 
physically valuable solutions out of a manifold of mathematically 
possible. One method is to add an artificial viscosity term to the right hand side of \eqref{pde_system}, to get the equation:
\begin{equation}
 \partial_t \mathbf{d} + \mathrm{div}_{\mathbf{x}} \, \mathbf{f}(\mathbf{d}) = \varepsilon \nu \Delta \mathbf{d} , 
\qquad \varepsilon > 0
\end{equation}
and then consider the limiting case $\varepsilon \rightarrow 0$. 
This idea is motivated by physical diffusion which broadens sincere discontinuities to differentiable steep gradients at the 
(microscopic) length  scale of the mean free path of the particles. The physical solution of the weakly formulated problem is thus the 
zero diffusion limit of the diffusive problem. However, in practice 
this limit is difficult to calculate analytically, 
and hence simpler conditions have to be found. 
A common technique to do this is motivated by continuum physics 
as well. Here an additional conservation law is set to hold for another 
quantity - the entropy of the fluid flow - as 
long as the solution remains smooth. Moreover, it is known that along 
admissible shocks 
this physical variable never decreases, and the conservation law for the 
entropy can be formulated as an inequality. 

We denote the (scalar valued) entropy function by $\sigma(\mathbf{d})$ and the 
entropy flux function by $\phi(\mathbf{d})$, and they satisfy
\begin{equation}  \label{cons_entro}
 \partial_t \sigma(\mathbf{d}) + \mathrm{div}_{\mathbf{x}} \, \phi(\mathbf{d}) = 0.
\end{equation} 
Assuming the functions to be differentiable, we may rewrite this conservation law 
via the chain rule and the equation \eqref{pde_system} as
\begin{equation}
 \nabla_{\mathbf{d}} \sigma(\mathbf{d})  \, \mathrm{div}_{\mathbf{x}} \, \mathbf{f}(\mathbf{d})   = \mathrm{div}_{\mathbf{x}} \, \phi (\mathbf{d}),
\end{equation}
where in higher-dimensional case the appearing matrices of gradients 
have to fulfillfurther 
constraints, see e.g.~\citep{godlewski+raviart_1992}. For scalar equations, it is 
always possible to find an entropy function of this kind. 
Furthermore it is assumed that the entropy function is convex, i.e.
\begin{equation} 
 \nabla_{\mathbf{d}}^2 \sigma > 0, \qquad \textrm{ for all }  \, \mathbf{d} \in U.
\end{equation}  
To get our actual entropy condition, we first rewrite our entropical conservation law \eqref{cons_entro} in the viscous form 
\begin{equation} \label{entro_visco}
 \partial_t \sigma(\mathbf{d}) + \mathrm{div}_{\mathbf{x}} \, \phi(\mathbf{d}) = 
\varepsilon \nabla_{\mathbf{d}} \sigma(\mathbf{d}) 
\Delta \mathbf{d}.
\end{equation}
Integrating over an arbitrary time interval $\left[ t_0, t_1 \right]$ and a connected set $\Omega \subset \mathbb{R}^n$, and using partial integration, we find that
\begin{equation}
\begin{aligned}
  & \int\limits_{t_0}^{t_1} \int\limits_{\Omega} \left( \partial_t \sigma(\mathbf{d}) + \mathrm{div}_{\mathbf{x}} \, \phi(\mathbf{d}) \right) \, \d V \, \d t \\
 = & \varepsilon \int\limits_{t_0}^{t_1} \int\limits_{\partial \Omega} \left( \nabla_{\mathbf{x}} \mathbf{d} \, \nabla_{\mathbf{d}} \sigma(\mathbf{d})\right) \cdot \mathbf{n} \, \d S \, \d t \\
 - &\varepsilon \int\limits_{t_0}^{t_1} \int\limits_{\Omega} \nabla_{\mathbf{x},i} {\mathbf{d}} \, \underbrace{\nabla^2 \sigma(\mathbf{d})}_{>0} \, \nabla_{\mathbf{x},i} \mathbf{d} \, \d V \, \d t.
\end{aligned}
\end{equation} 
When we now consider our non-diffusive limit $\varepsilon \to 0$, the 
first term on the right-hand side vanishes without further restriction 
whereas the second term has to remain nonpositive. With partial 
integration and divergence theorem we get our entropy condition
\begin{equation} \label{entropy_condition}
\begin{aligned}
 \int\limits_{\Omega} \sigma(\mathbf{d}(\mathbf{x},t_1)) \, \d V  	& \leq  \int\limits_{\Omega} \sigma(\mathbf{d}(\mathbf{x},t_0)) \, dV \\
  & - \int\limits_{t_0}^{t_1} \int\limits_{\partial \Omega} \phi(\mathbf{d}) \cdot \mathbf{n} \, \d S \, \d t. \\
\end{aligned}
\end{equation}
For bounded, continuous pointwise solutions 
$\mathbf{d}^*$ of (\ref{entro_visco}) such that $\mathbf{d}^* \to \mathbf{d}$ for $\varepsilon \to 0$, the 
vanishing viscosity solution $\mathbf{d}$ is a weak solution of the initial 
value problem (\ref{cauchy_problem_integral}) and fulfills entropy condition 
(\ref{entropy_condition}). Generally spoken, applying the entropy condition 
to systems with shock solutions unveils those propagation velocities that ensure 
that no characteristics rise from discontinuities which would be 
non-physical. For detailed motivation, stringent argumentation
and proofs to mathematical techniques presented 
in this section we refer to \citep{harten+hyman+lax_1967} respectively 
to~\citep{leveque_1991}. 

\subsection{Numerical Artificial Viscosity} 
\label{Artificial_Viscosity_and_Flux}

As mentioned we are looking for high-resolution 
methods for nonlinear PDEs derived from hyperbolic conservation laws. 
In the past decades major efforts have been made in 
developing numerical methods for these problems 
that are at least of second order. 
One patent attempt to finding such a high-resolution method is to adapt 
a well-known high-order method for linear 
problems for nonlinear problems (such as the Lax-Wendroff scheme~\citep{lax+wendroff_1960}). 

As illustrated above we can add an 
artificial viscosity term to the conservation law in a way that 
the entropy condition is satisfied and non-physical solutions 
are excluded. We are keen to design this 
viscosity in such a manner that it affects sincere discontinuities but vanishes 
sufficiently elsewhere so that the order of accuracy can be maintained in those 
regimes where the solution is smooth. The idea of numerical artificial viscosity was inspired by physical dissipation 
mechanisms and dates back 
more than half a century to~\citep{vonneumann+richtmyer_1950}.

We denote as customary the approximate solution of the exact density
$\mathbf{d}(x,t)$ at discrete grid points $\mathbf{d}(x_j,t_n)$
by $\mathbf{D}^n_j$, and set 
$\mathbf{D} =  \left [ \mathbf{D}_1 \ldots \mathbf{D}_k \right ]^{\mathrm{T}}$,
where $k$ is the total number of grid points. The numerical 
representation of the flux function $\mathbf{f}(\mathbf{d})$ is denoted 
respectively by $\mathbf{F}(\mathbf{D})$, where $\left[ \mathbf{F}(\mathbf{D}) \right]_j = \mathbf{f}(\mathbf{D}_j)$. 
The numerical flux function gets modified by a an artificial viscosity 
$\mathbf{Q} \left[ \mathbf{D} \right]_j$ for instance in the following way:
\begin{equation}
 \left[ \mathbf{F}_{\text{visc}} ( \mathbf{D} ) \right]_j = \left[ \mathbf{F} 
( \mathbf{D}) \right]_j - h  \left[ \mathbf{Q} \big( \mathbf{D} \big) \right]_j \big( 
\mathbf{D}_{j+1} - \mathbf{D}_j \big).
\end{equation}
where $h$ denotes the size of the spatial discretization.
Since the original design of this additional viscous 
pressure in the scalar form $Q = c_2 \rho (\Delta \mathbf{u})^2$, $c_2 \in
\mathbb{R}$ as suggested in~\citep{vonneumann+richtmyer_1950} for one dimensional 
advection $\partial_t \mathbf{d} + a 
\partial_x \mathbf{d} = Q \partial_{xx} \mathbf{d}$, it has undergone a number 
of modifications and generalizations. It has turned out to be numerically 
preferable to add a linear term (see~\citep{landshoff_1955}) in order to control 
oscillations. Generalizations to multi-dimensional flows mostly 
retain the original analogy to physical dissipation 
and reformulate the velocity term accordingly, see e.g.~\citep{wilkins_1980}. 

The artificial viscosity 
broadens shocks to steep gradients at some characteristic length scale, but should not cause too large smearing. 
The concrete composition and implementation of this artificial viscosity 
coefficient $\mathbf{Q}$ depends on the application. 
As an example we discuss the following form of the tensor 
of the artificial viscosity in higher-dimensional RHD numerics.  
Similar forms of artificial viscosity can be found also in 
pure hydrodynamics and MHD calculations in 2D and 3D. 
 
Tscharnuter and Winkler~\citep{tscharnuter+winkler_1979} have pointed out 
that the viscous pressure in 3D radiation hydrodynamics 
has to unravel normal stress, quantified by the 
divergence of the velocity field and shear stress, which 
is expressed by the symmetrized gradient of 
the velocity field according to the general theory of viscosity. 
It is designed to switch on only in case of compression ($\mathrm{div}_{\mathbf{x}}\, \mathbf{u} < 0$),
and this is all ensured by the form
\begin{equation} \label{Q_Tscharnuter}
 \mathbf{Q} = 
- q_2^2 l_{\text{visc}}^2
\rho \max(-\mathrm{div}_{\mathbf{x}} \, \mathbf{u}, 0)  \left( \Big[
\nabla \mathbf{u} \Big]_s  - \frac{1}{3}  \mathbf{e} \, 
\mathrm{div}_{\mathbf{x}} \, \mathbf{u}
\right),
\end{equation} 
where the symmetrization rule is defined componentwise for the lower indices as
$$
{\left([\nabla \mathbf{u}]_s\right)}_{ij} = 
\frac{1}{2}(\nabla_i u_j + \nabla_j u_i).
$$

\section{Numerical Artificial Viscosity in Curvilinear Coordinates} \label{Artificial_Viscosity}

We introduced artificial viscosity in form of a three dimensional 
viscous pressure tensor \eqref{Q_Tscharnuter} in section \ref{Artificial_Viscosity_and_Flux} along
the lines of~\citep{tscharnuter+winkler_1979}. In this section we want to point out, how
such a definition must be adapted for curvilinear coordinates in 
order to ensure tensor analytical consistency. 

When formulating PDEs derived from  
hyperbolic conservation laws on a curvilinear grid,
the tensorial equations \eqref{cauchy_problem_integral} have to be transformed 
to the according coordinate system. Not only the vectorial 
and tensorial quantities have to be transformed but also the differentiation 
operators, in particular the divergence operator in our case. The appropriate 
framework to do this is provided by differential geometry. 
Like the gradient of a scalar is natively a covector, there are 
rules for co- and contravariant indices of tensors such as the one we 
are interested in. The crucial term in 
(\ref{Q_Tscharnuter}) is the symmetrized velocity gradient $\Big[ \nabla \mathbf{u} \Big]_s$ that accounts 
for shear stresses, and one sees that the form \eqref{Q_Tscharnuter} comes into conflict 
with the demand of vanishing trace $(\mathrm{Tr} \mathbf{Q} = 0)$ when 
the divergence term is simply of the form 
$ \,\, \mathbf{e} \, 
\mathrm{div}_{\mathbf{x}} \, \mathbf{u} \,\,$, as we find it commonly in several  
MHD and RHD grid codes. 


\textbf{Proposition:} The correct form of the viscous pressure tensor \eqref{Q_Tscharnuter}
in general coordinates is 
\begin{equation} \label{art_visc_corr}
 \mathbf{Q} = - 
q_2^2 l_{\text{visc}}^2 \rho \max(-\mathrm{div}_{\mathbf{x}} \,  \mathbf{u}, 0)  
\left( \Big[ \nabla \mathbf{u} \Big]_s - \frac{1}{3} \mathbf{g} \, \mathrm{div}_{\mathbf{x}}
\,  \mathbf{u} \right) .
\end{equation}
We show that this tensor has the desired properties. 
The viscous pressure tensor must be symmetric by definition, 
i.e. $Q_{ij}=Q_{ji}$, which can be easily verified from \eqref{art_visc_corr}.
Also, the trace of the tensor has to vanish $\left(\mathrm{Tr} \mathbf{Q} = {Q^i}_i = 0 \right)$, 
as pointed out in~\citep{vonneumann+richtmyer_1950}. To show that this holds for \eqref{art_visc_corr}, 
we first consider its {\sl native} covariant components 
\begin{equation} \label{art_vis_corr_comp}
 Q_{ij} =  - \mu_2 
\max(- \mathrm{div}_{\mathbf{x}} \, \mathbf{u}, 0)
 \left( \frac{1}{2}
(\nabla_i u_j + \nabla_j u_i ) - \frac{1}{3} g_{ij} \, \mathrm{div}_{\mathbf{x}}
\,  \mathbf{u} \right),
\end{equation}
where we have renamed $q_2^2 l_{\text{visc}}^2 \rho = \mu_2$. Next, we need to raise an index with the metric,
\begin{align*}
 {Q^i}_j &= Q_{lj}g^{li} \\
      &= - \mu_2 \max(- \mathrm{div}_{\mathbf{x}} \, \mathbf{u}, 0) \left( \frac{1}{2} g^{li}(\nabla_l u_j + \nabla_j u_l ) - 
    \frac{1}{3} {\delta^i}_j \, \mathrm{div}_{\mathbf{x}} \,  \mathbf{u} \right),
\end{align*}
and use the essential identity $g^{li}g_{lj}={g^i}_j ={\delta^i}_j$. 
The Ricci Lemma $\nabla_i g_{jk} = \partial_i g_{jk} - {\Gamma^l}_{ij} g_{lk} - 
{\Gamma^l}_{ik} g_{jl} = 0$ 
for the fundamental tensor naturally also holds for the 
contravariant components and we can permute $g$ into the derivatives 
$\nabla_l g^{li} u_j$ and $\nabla_j g^{li} u_l$ which yields 
twice the divergence $\nabla_i u^i= \mathrm{div}_{\mathbf{x}} \,
\mathbf{u}$ when we conduct the contraction $j \to i$. 
In three dimensions the summation ${\delta^i}_i=3$ and we obtain 
our desired result 
\begin{equation*}
 {Q^i}_i = \ldots = \left( \frac{1}{2} (2 \, \mathrm{div}_{\mathbf{x}} \, \mathbf{u}) - \frac{1}{3}3 \, \mathrm{div}_{\mathbf{x}} \, \mathbf{u} \right) = 0.
\end{equation*}
The commonly used (see e.g.~\citep{dorfi_1999} in RHD, 
~\citep{iwakami+al_2008} in MHD, ~\citep{fryxel+al_2000} in MHD)
 form of $\mathbf{Q}$ \eqref{Q_Tscharnuter} is not compatible 
with these requirements since the symmetrization is only 
defined for lower indices, whereas the unit tensor $\mathbf{e}$ of a metric 
space is only defined for mixed indices, meaning there is no 
such thing as $\delta_{ij}$. 
However, the above mentioned and other authors 
such as~\citep{mihalas+mihalas_1984} have neglected that little 
inconsistency since they have considered mixed indices from the start
respectively cartesian or affine coordinates. Nonlinear 
corrections have been suggested by~\citep{benson+schoenfeld_1993} 
albeit they do not explicitly concern curvilinear coordinates and
are based on a TVD approach. 

In several hydro- and MHD-codes that include non-Cartesian grids 
such as Pluto~\citep{mignone+al_2007}, the geometric source terms are coded 
explicitly for several geometries (polar, cylindrical, spherical), and not 
only for the artificial viscosity flux. The suggestions made e.g. 
by~\citep{vinokur_1974} lead to geometrical source terms that 
correct curvilinear grid effects. However the strong conservation form 
as elaborated by~\citep{warsi_1981} 
would need to appeal to our differential 
geometrically consistent approach in order to deal with the viscosity in an 
intrinsically consistent way. Especially when the metric tensor itself is 
not only a function of space but also time-dependent (as discussed in Section
\ref{Intro}), the latter approach reaches its limits. Our correction affects curvilinear coordinates in multiple dimensions, 
whereas it is not necessary that the coordinates are orthogonal respectively the metric 
tensor does not need to be diagonal. Our initial motivation to study 
more general coordinates comes from the idea to 
generate problem-oriented coordinate systems for astrophysical 
numerical calculations. In a following paper we want to present some 
feasible approaches to grid generation under certain physical 
restrictions. Such nonlinear grids that are adaptive in multiple dimensions 
have time-dependent metric tensors and thus benefit directly from 
our consistent definition. On the contrast, when using adaptive mesh refinement, 
the metric tensor remains geometrically constant in time. 


In order to support the theoretical results in this work, in the upcoming section we study as an example a very simple 
velocity field with non-vanishing divergence and visualize the according artificial viscosities for the two presented cases. 

\section{Application and Visualization}

The most common application of curvilinear coordinates in 3D is the 
map $(x,y,z) \to (r \in \mathbb{R}^+, \, \vartheta \in [0, \pi], \, \varphi \in [0,2\pi])$ with 
$x = r \sin \vartheta \cos \varphi$, $y = r \sin \vartheta \sin \varphi$ 
and $z = r \cos \vartheta$ as spherical coordinates. The corresponding diagonal
covariant metric components in this simple orthogonal case 
are given by $\mathrm{diag} \, (1, r, r\sin\vartheta)$. 

\subsection{Toy Model Velocity Field}

As the presented considerations for artificial viscosity on nonsteady 
curvilinear coordinates originate from astrophysical applications, we
want to consider a velocity field with a certain practice in RHD. 
We study a toy model of a self-similar fluid flow solution, namely the velocity field given by 
\begin{equation} \label{eq:toyfield} 
 \mathbf{u}_{\text{Ex}} = \frac{\mathbf{x}}{\sqrt{x^2+y^2+z^2}} = \frac{\mathbf{x}}{r}
\end{equation}
here in Cartesian coordinates. Such self-similar solutions appear in idealized spherical models of 
stars for example as shocks driven by radial stellar pulsations.  

This vector field is obviously symmetric with respect to the origin and has
a non-vanishing divergence $\mathrm{div}_{\mathbf{x}} \, \mathbf{u}_{\text{Ex}} = 2/r$. 
The covariant components of this vector field are given in any other coordinates by scalar product 
with the base vectors, i.e. ${u}_{\text{Ex},i} = \mathbf{u}_{\text{Ex}} \cdot \mathbf{e}_i$. 
This leads to the covariant components $(1,0,0)$ in spherical coordinates. 
%
The nonzero covariant components of the tensorial part $\,( [\nabla \mathbf{u} ]_s  - \frac{1}{3}  \mathbf{e} \,  
\mathrm{div}_{\mathbf{x}} \, \mathbf{u} ) \,$ of the artificial viscosity \eqref{art_visc_corr} are given for this field by
\begin{equation}\label{numericalEqA}
 Q_{rr} = -\frac{2}{3r}, \quad Q_{\theta \theta} = \frac{r}{3}, 
\quad Q_{\phi \phi} = \frac{r \sin^2 \theta}{3}.
\end{equation}
In the following section we visualize the tensor of artificial viscosity (TAV) for the velocity 
field \eqref{eq:toyfield}. A uniform distribution of the leading eigenvalues
over the whole domain is expected due to the symmetry of the vector field. 
%

One easily verifies the identity ${Q^i}_i=0$ summing over the 
mixed components ${Q^r}_r = -2/3r, \,\, {Q^{\theta}}_{\theta}= 1/3r, \,\, 
{Q^{\phi}}_{\phi}= 1/3r$. 

With the previous version of the TAV from~\citep{tscharnuter+winkler_1979} we 
would get the following covariant components of the artificial viscosity tensor.
\begin{equation}\label{numericalEqB}
  Q_{rr} = -\frac{2}{3r}, \quad Q_{\theta \theta} =-\frac{2}{3r} + r, 
\quad Q_{\phi \phi} = -\frac{2}{3r} + r \sin^2 \theta.
\end{equation}
The visualization of this non-metric version of the artificial 
viscosity for the symmetric velocity field \eqref{eq:toyfield} shows obviously a field unequal in strength and direction over the 
whole domain. In a numerical calculation this will clearly lead to 
artificial anisotropies in the flux of the density field and
destroy all efforts in constructing a higher-order conservative numerical 
scheme with artificial viscosity.  
%

\subsection{Visualization of scalar and tensor fields} \label{vish}

We can see the incorrect vs. correct behavior immediately by
even just displaying the major eigenvalue or trace as one indicator. 
However, since the major eigenvalue represents only one
degree of freedom out of the six available in the tensor field, a
technique depicting all six components is a more objective
way for validation.

\begin{figure}
  \centering
  \subfigure[][$\lambda_{\max}$]{\includegraphics[width=\hsize]{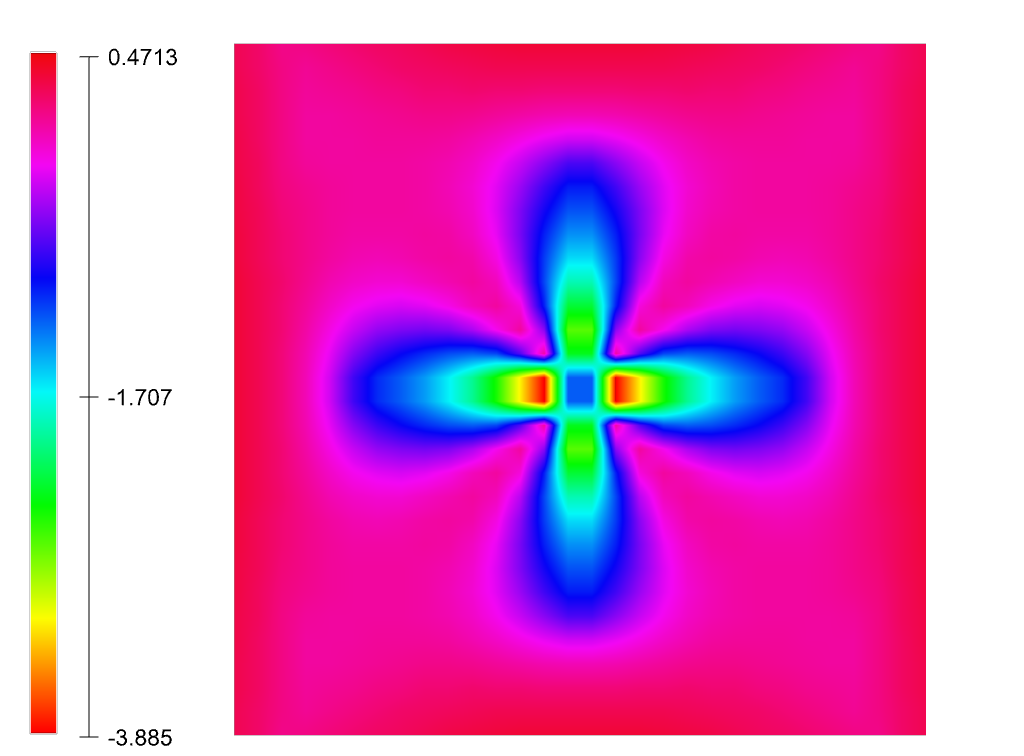} }
  \subfigure[][$\lambda_{\max}$]{\includegraphics[width=\hsize]{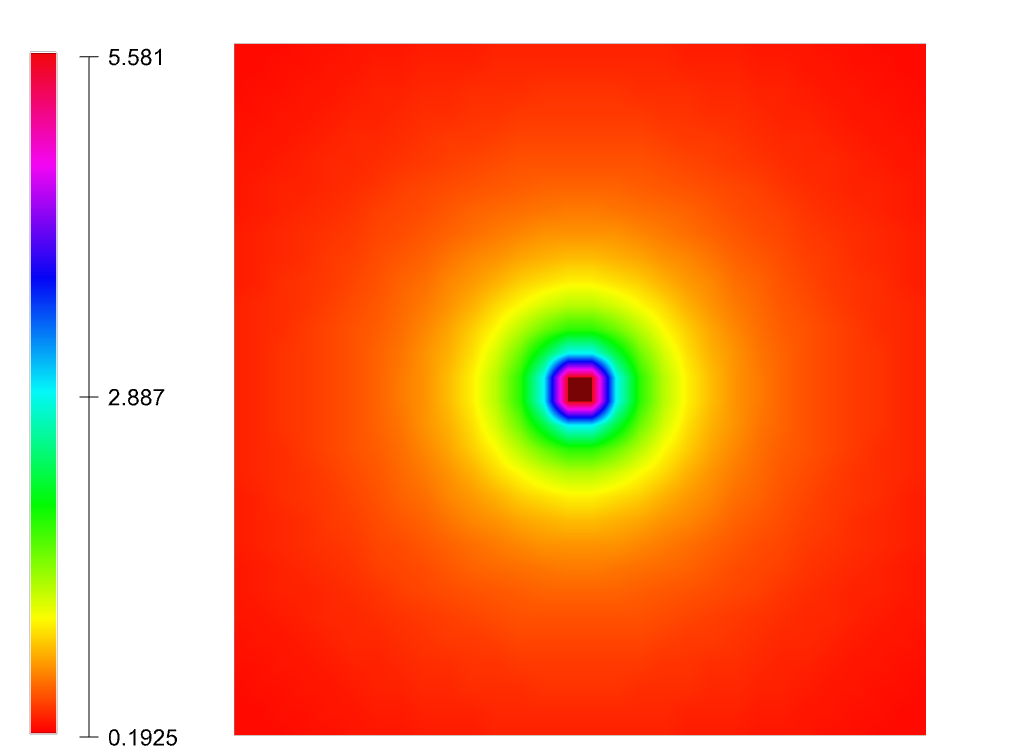}} 
  \caption{Major eigenvalue the incorrect (left)
  and correct (right row) viscosity, shown along the XZ plane.
  } 
  \label{EigenValues}
\end{figure}

\begin{figure}[htb]
\centering
\subfigure[][$\mathrm{Tr} \mathbf{Q}_{\text{incorr}}$]{\includegraphics[width=\hsize]{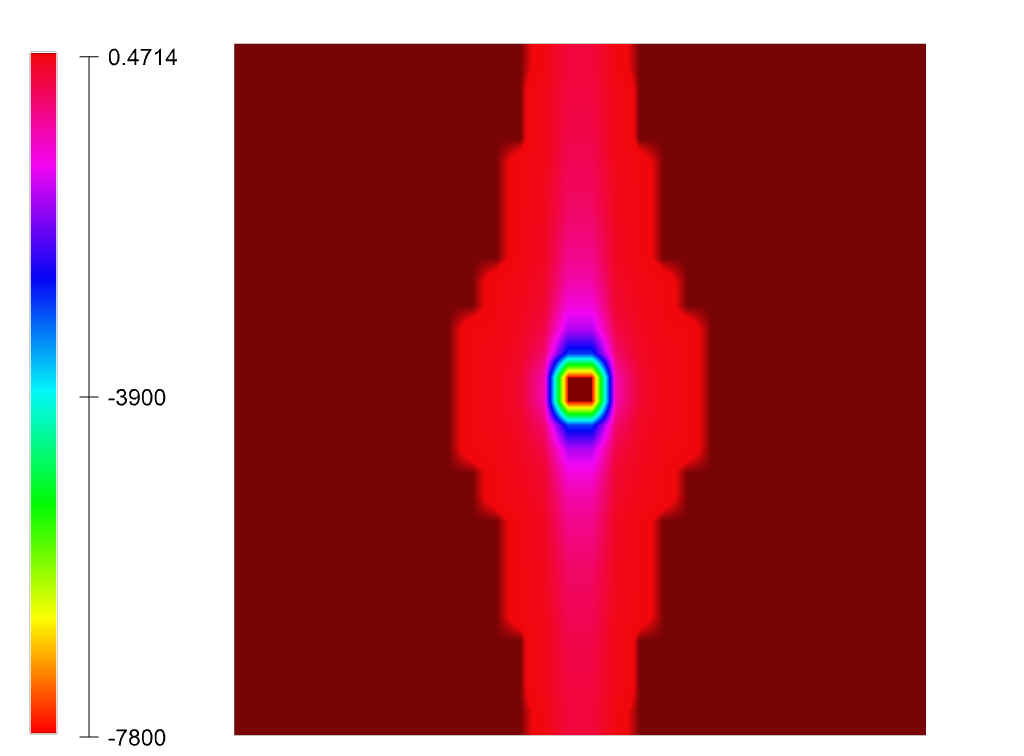} \label{trace_incorrect}}
\subfigure[][$\mathrm{Tr} \mathbf{Q}_{\text{corr}}$  ]{\includegraphics[width=\hsize]{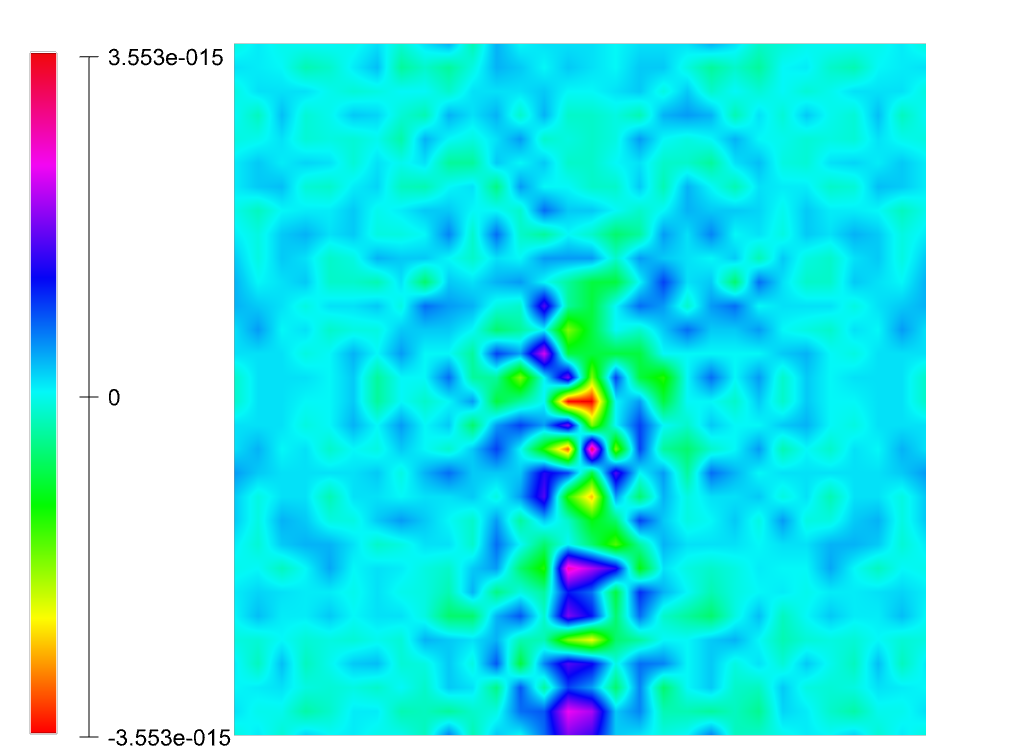}\label{trace_correct}  } 
\caption{Trace the incorrect (left)
and correct (right row) viscosity, shown along the XZ plane.
Note the range of values in Fig.~\ref{trace_correct} - what we see is just
numerical noise, the trace is perfectly zero within numerical precision.
} 
\label{TraceViz}
\end{figure}

\begin{figure}[htb]
\centering
\subfigure[][Incorrect]{\includegraphics[width=\hsize]{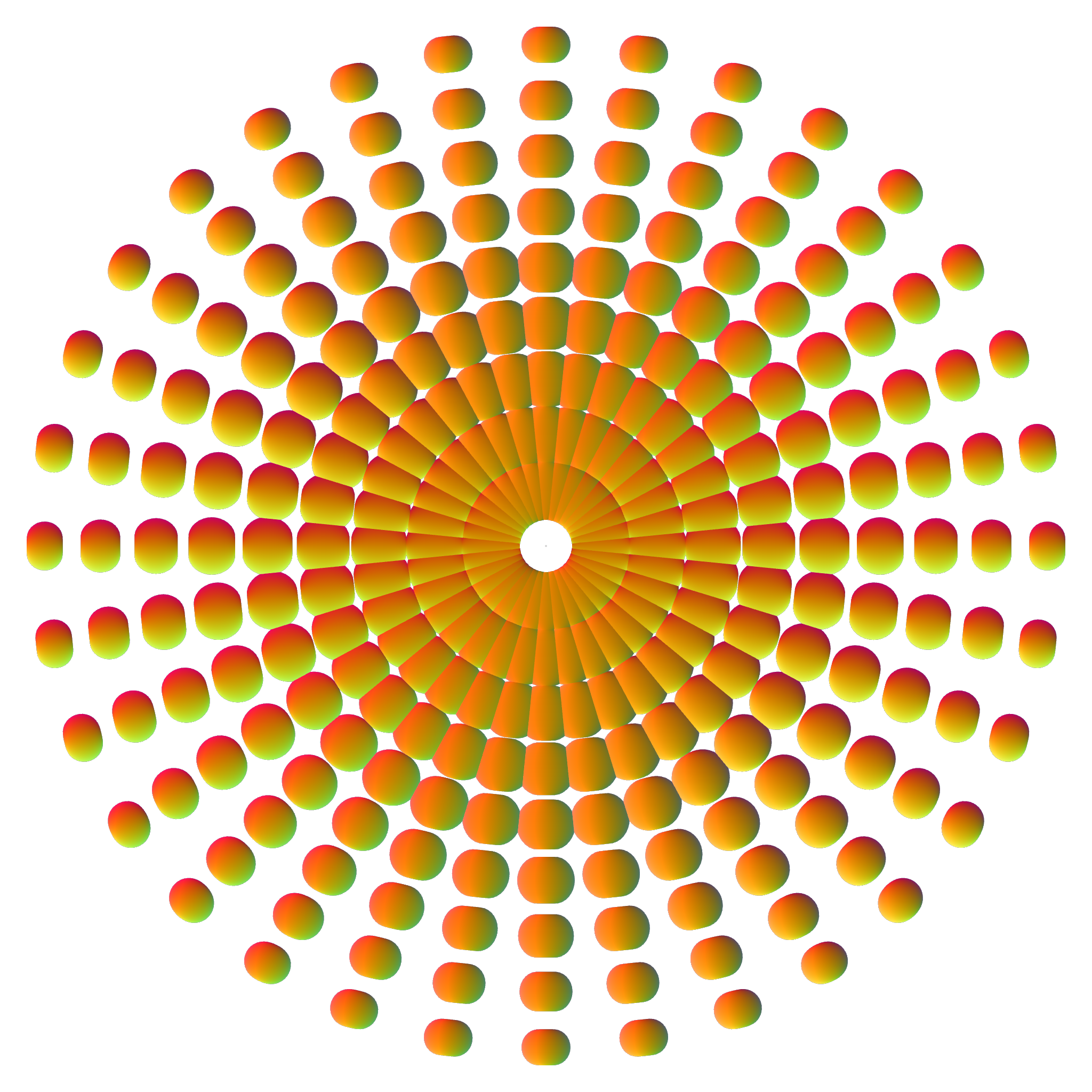}  \label{ReynoldsOld} }
\subfigure[][Correct]{\includegraphics[width=\hsize]{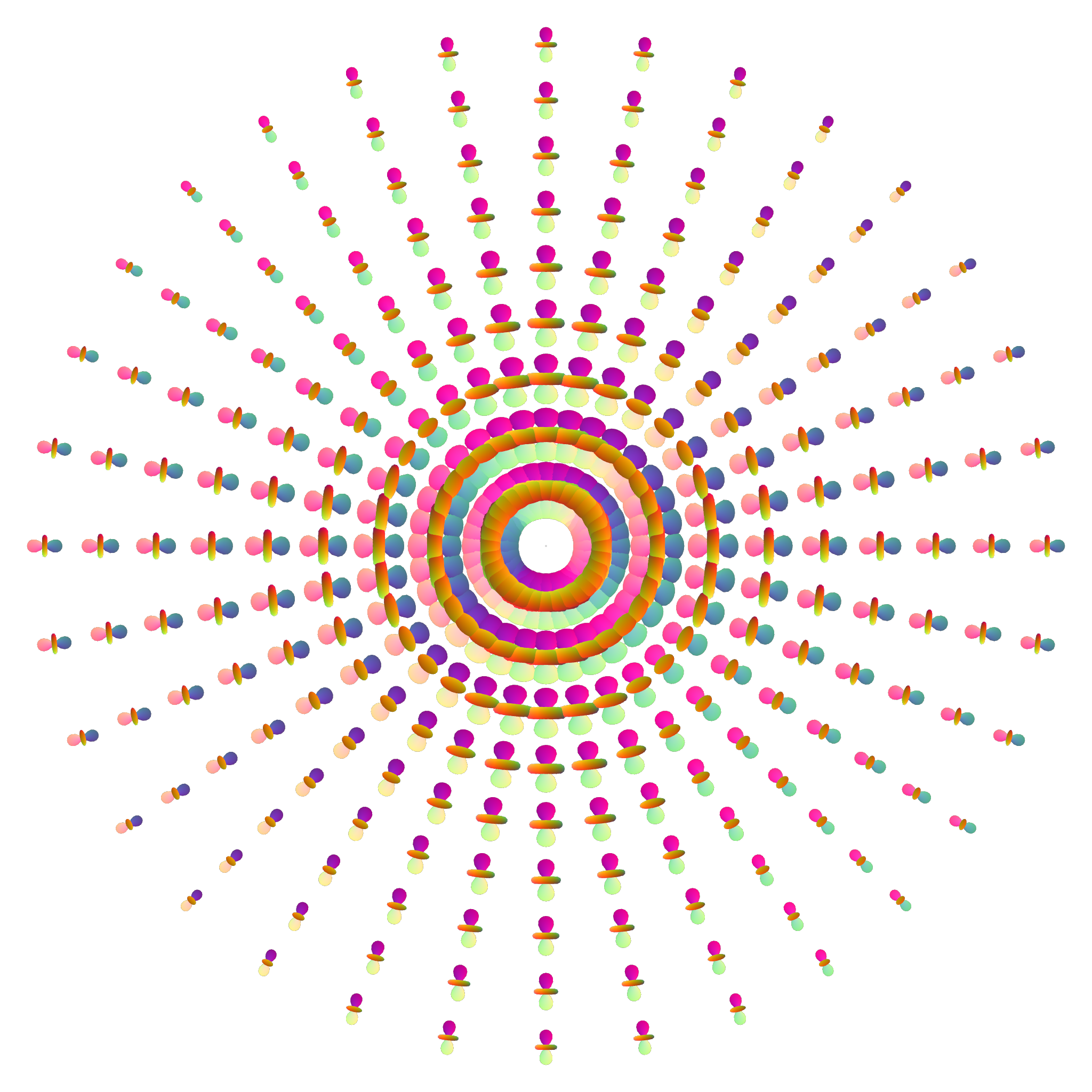}  \label{ReynoldsNew} } \\
\caption{Reynold glyphs~\citep{reynoldglyph} of the incorrect and correct viscosity tensor.
The incorrect tensor showing a strongly negative component, whereas the correct
tensor is balanced, indicating zero trace.
} 
\label{ReynoldsTensorViz}
\end{figure}

We used the Vish Visualization Shell~\citep{VISH06} to  numerically 
sample \eqref{numericalEqA} and \eqref{numericalEqB} on 
a uniform grid for analyzing scalar fields (Fig.~\ref{EigenValues}, Fig.~\ref{TraceViz} ) 
and a radial sampling distribution (Fig. \ref{ReynoldsTensorViz}) for the
full tensor field.

Fig.~\ref{EigenValues} displays a structure of the eigenvalue corresponding to the
major eigenvector of the TAV on the XZ plane (i.e., in the plane $y=0$), evidently
showing some asymmetric `funny' coordinate-dependent
behavior of the incorrect TAV, whereas the 
correct TAV is radially symmetric, as desirable.
The tensor field is symmetric and of rank two, however it is not
positive definite and may exhibit vanishing trace.
The correct TAV is trace-free in the entire domain, whereas the trace
of the incorrect TAV ranges through positive values (for large radial
distances) to large negative values close to the coordinate origin,
as depicted in Fig.~\ref{TraceViz} .
A direct visualization method depicting the full six-dimensional degrees of freedom the 
tensor field is thus favorable, or even rather essential.
However, many direct visualization methods for tensor fields require positive
definiteness and are thus not applicable to this data.
Only very few methods are suitable for general tensors.
Fig.~\ref{ReynoldsTensorViz} shows so-called Reynold glyphs~\citep{reynoldglyph}
for the TAV field.
A Reynold glyph is the surface generated by mapping a tangential vector $v(P)$ at
each data sample point $P$ as
\begin{equation*} 
  P\rightarrow Q\left(v(P), v(P)\right).
\end{equation*}
Such glyphs shown at each sampling point provide a direct visualization of the 
full six-dimensional degrees of freedom of the tensor properties. 
Reynold glyphs are able to also depict negative definite tensors, whereas a quadric surface 
(ellipsoids representing $P\rightarrow 1/\sqrt{Q\left(v(P), v(P)\right)}$ becomes hyperbolic for 
negative eigenvalues and problematic for visualization purposes.
The Reynold glyph directly show the ``directional value'' $Q\left(v(P), v(P)\right)$ of the
tensor field $Q$ in the direction $v$ around a the sampling point $P$ - the resulting 
surface is intersecting the sampling point $P$ whenever $Q\left(v(P), v(P)\right)=0$,
which is the case for points where the tensor is degenerating and not positive definite.
In such areas the glyph will visually appear like two intersecting surfaces
corresponding to the isopotential surfaces of second order spherical harmonic functions.
These both surface components represent positive and negative eigenvalues of the
tensor field - if both positive and negative component ``counter-balance'' themselves
they therefore indicate vanishing trace, which is the sum of the eigenvalues. If only
one surface component is visible, then the tensor is either positive or negative definite
on that certain point.

\begin{figure}[htb]
\centering
\subfigure[][Detail of the incorrect viscosity]{\includegraphics[width=\hsize]{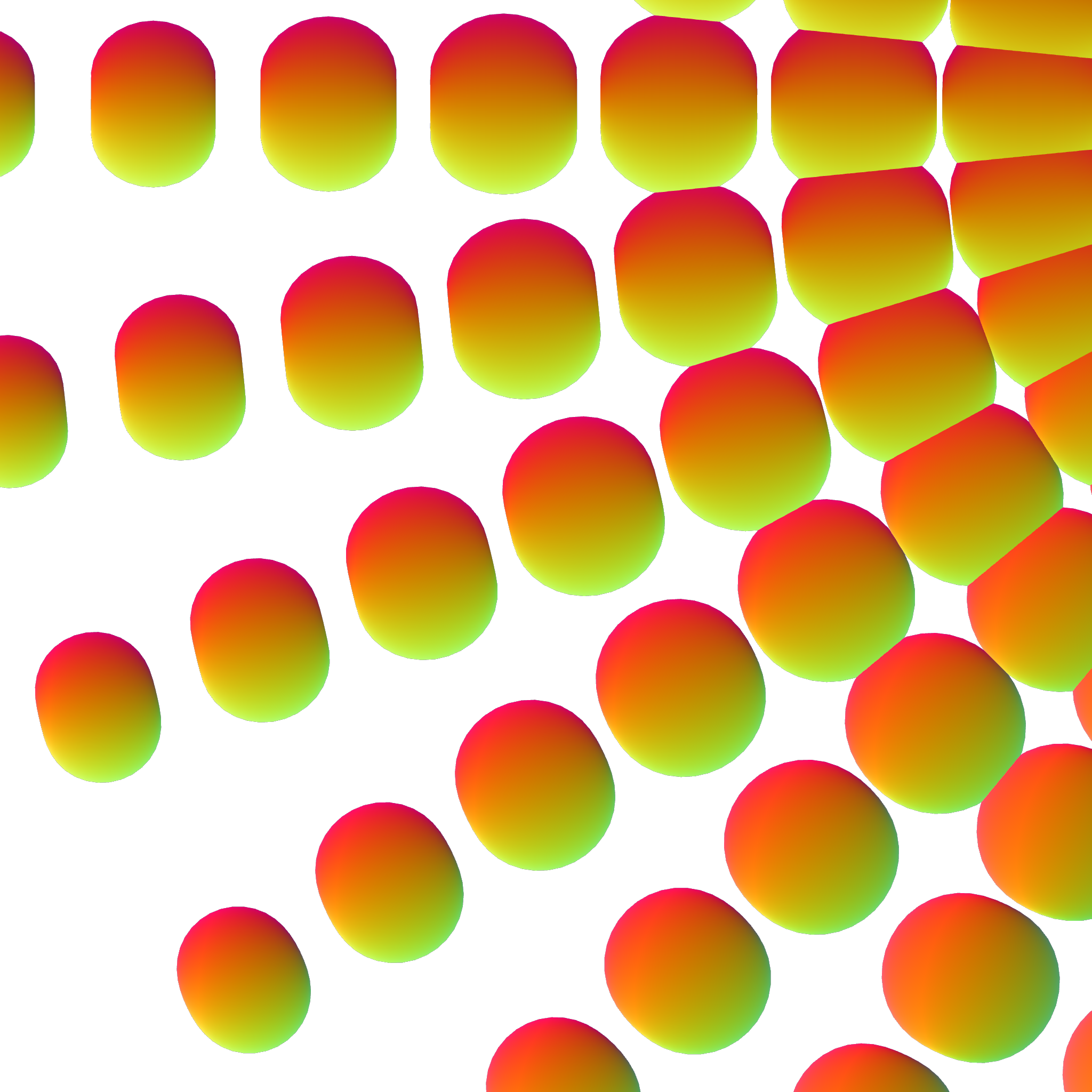} \label{ReynoldsDetailOld}  }
~
\subfigure[][Detail of the correct viscosity]{\includegraphics[width=\hsize]{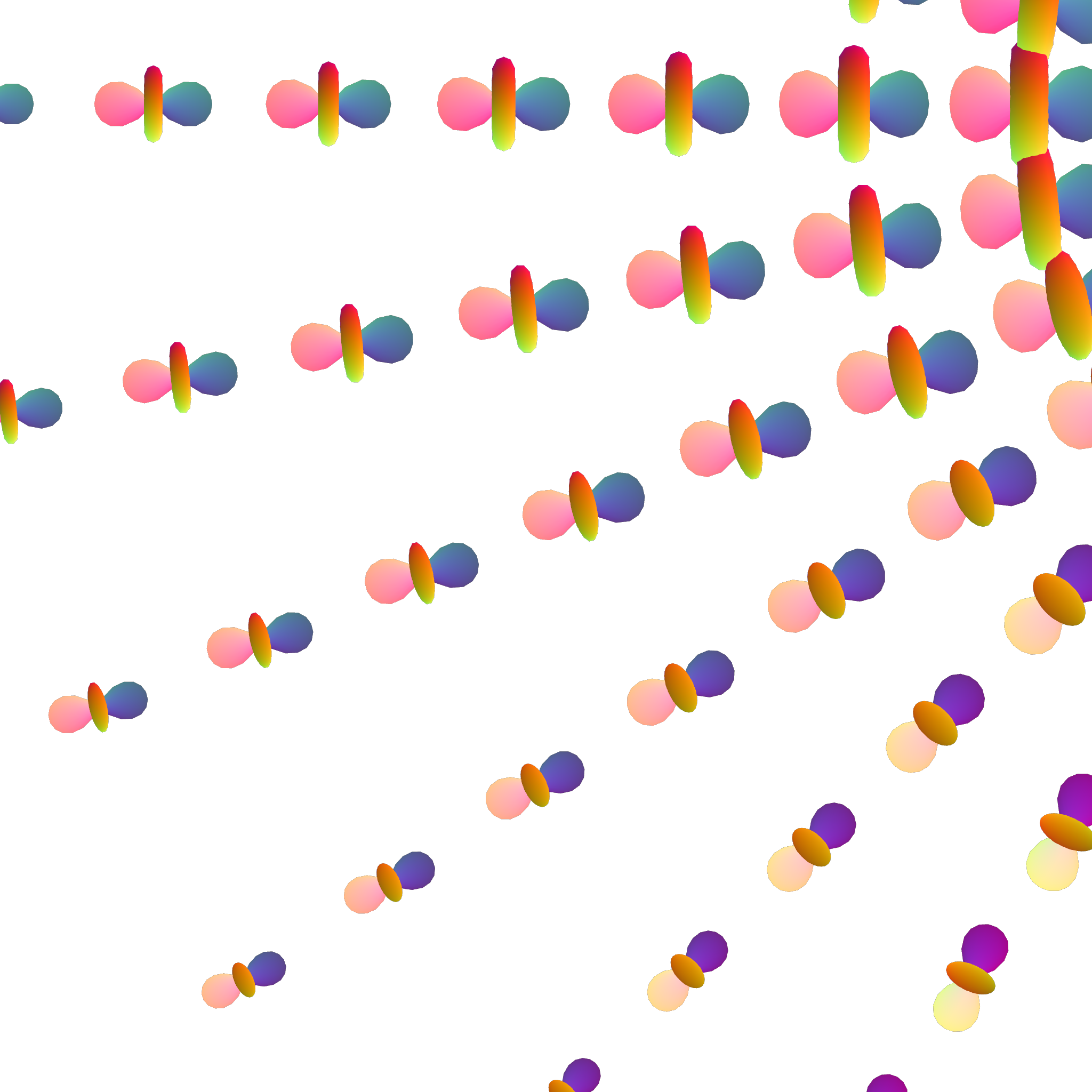}  \label{ReynoldsDetailNew} }\\
\caption{Detailed Reynold glyphs~\citep{reynoldglyph} of the incorrect and correct
viscosity tensor.
Glyphs of the correct tensor show spherical harmonics that
are balanced in their positive and negative half-widths.
} 
\label{ReynoldsTensorViz2}
\end{figure}

We used a radial sampling for the direct visualization of the tensor field in order to
minimize coordinate artificats.
As depicted in Fig.~\ref{ReynoldsNew} and Fig.~\ref{ReynoldsDetailNew} for the correct TAV all ``modes'' are equivalently
represented, indicating vanishing trace of the tensor field while being radially aligned with the
underlying coordinate system.
In contrast, the incorrect TAV, Fig.~\ref{ReynoldsOld} and Fig.~\ref{ReynoldsDetailOld},  exhibits a dominantly negative trace.

%
\section{Conclusions}

We studied a generalization of the tensor of numerical artificial viscosity for 
curvilinear coordinates and compared our result to previous definitions found in literature. 
We analyzed a symmetric toy velocity field and visualized its viscosities. 
Clearly, the non-metric version of the TAV as used by many authors of 
hydro- (HD) respectively magneto-hydro- (MHD) or radiation hydrodynamic-codes
(RHD) leads to incorrect results in curvilinear coordinates, whereas our suggestion for 
the numerical artificial viscosity gives geometrically consistent results.

\section{Appendix} \label{appendix}

As mentioned in section \ref{Intro}, the benefit of the strong formulation 
of Cauchy problems arising from physical conservation laws will be discussed in more detail here.

Following the ideas of ~\citep{warsi_1981}, in non-steady coordinates the geometrically 
consistent strong form of tensorial conservation laws \eqref{erh} is given as
\begin{equation} \label{cauchy_cons}
 \partial_t \left( \sqrt{|\mathbf{g}|} \, \mathbf{d} \right) + 
\partial_i \left( \sqrt{|\mathbf{g}|} \, \mathbf{f}(\mathbf{d}) 
\cdot \mathbf{{e}}^i \right) = 0, 
\end{equation}
where $\mathbf{d}$ and $\mathbf{f}$ are to be decomposed according to their native tensorial components. The scalar multiplication with 
the $i$-th contravariant base vectors $\mathbf{{e}}^i$ yields a projection on the contravariant 
coordinate lines which in case of generalized grids can differ in direction and length of their covariant counterparts. 
Equation \eqref{cauchy_cons} gives also the integral form of the conservation law, which should be 
treated numerically in a correct way for non-steady coordinates in any finite-volume discretization. 
The important difference to the componentwise 
structure $\nabla (.) = \partial (.) + \Gamma (.)$, where Christoffel symbols account for the 
geometry is that in this case undifferentiated terms arise (see ~\citep{vinokur_1974}) which act 
like geometric sources in the equations and destroy conservativeness. A comprehensive proof of 
Vinokurs theorem using differential forms can be found in ~\citep{bridges_2008}.  

With non-steady curvilinear grids not only the nonlinearity of the metric tensor but also 
its time-dependence has to be taken into account numerically. The motion of the grid itself 
and its implications on the formulation of the set of equations respectively the 
calculation of the occurring fluxes is discussed in the following section. 

\subsection{Adaptive Grids} \label{Adaptive Grids}	

In fluid dynamics we distinguish two main reference systems that suit unequally 
for various applications. The Eulerian frame is the fixed reference system of an external 
observer in which the fluid moves with velocity $\mathbf{u}$ whereas the 
Lagrangian approach describes the physics in the rest frame of the fluid. 
Between these two systems, the transformation of an advection term for a density $\mathbf{d}$
(that moves with a relative velocity $\mathbf{u}$) is given via the material derivative
$D_t \mathbf{d}  = \partial_t \mathbf{d} + \mathbf{u}\cdot\nabla \mathbf{d}$.

Hence, when we work with comoving frames, the coordinate system respectively the computational 
grid is time-dependent. There is a number of purposes where strict Eulerian or Lagrangian 
grids are suboptimal and thus we need to consider the generalized the concept of the comoving frames.

We want to evolve the strong conservation form for time dependent general coordinate systems. 
The time derivative of a density $\mathbf{d}$ in the coordinate system
$\Sigma_{(\beta)}$ relative to a (e.g. static) coordinate system
$\Sigma_{(\alpha)}$ is given by ${\partial_t \mathbf{d}}_{(\beta)} = 
{\partial _t\mathbf{d}}_{(\alpha)} + 
\mathbf{\nabla}_{(\alpha)}\mathbf{d} \, {\partial_t \mathbf{x}}_{(\beta)}$
and from the view point of system $\Sigma_{(\alpha)}$, the time derivative 
is given as   
\begin{equation}
 {\partial_t \mathbf{d}}_{(\alpha)} = {\partial _t\mathbf{d}}_{(\beta)} - 
\dot{\mathbf{x}} \cdot \mathbf{\nabla}_{(\alpha)}\mathbf{d}^{\mathrm{T}},
\end{equation}
where $\dot{\mathbf{x}}$ denotes the grid velocity. The 
second term on the right side we call grid advection, see e.g.
~\citep{warsi_1981} and ~\citep{thompson+warsi_1985}. 
An inhomogeneous advective term of a conservation law 
$\mathbf{K} = \mathbf{d}_t + \mathrm{div} \, (\mathbf{u}^{\mathrm{T}} \mathbf{d})$
(in a fixed coordinate system) is given in the case of moving grid as
\begin{equation}
 \mathbf{K} = \mathbf{d}_t - \dot{\mathbf{x}} \cdot 
\mathbf{\nabla}\mathbf{d}^{\mathrm{T}} + \mathrm{div} \,
(\mathbf{u}^{\mathrm{T}}\mathbf{d}). 
\end{equation}
When we apply the appropriate transformation prescriptions to the spatial derivatives, we 
gain the following form
\begin{equation} \label{conv_cons}
 \mathbf{K} = \mathbf{d}_t - \dot{\mathbf{x}} \cdot \frac{1}
{\sqrt{|\mathbf{g}|}}\partial_{i}
\left(
\sqrt{|\mathbf{g}|} \mathbf{{e}}^{i} \mathbf{d} \right) + 
\frac{1}{\sqrt{|\mathbf{g}|}}\partial_{i} \left(
\sqrt{|\mathbf{g}|} \mathbf{u} \cdot \mathbf{{e}}^{i} \mathbf{d} \right).
\end{equation}
Now this is not yet geometrically conservative, since it is not of 
an integral structure. Following the idea of the Reynolds transport theorem 
we consider the temporal derivative of the volume respectively 
the determinant of the time dependent metric tensor 
$\sqrt{|\mathbf{g(\mathbf{x},t)}|}$ in order to study the conservation of a density 
function in variable volumes and obtain the strong conservation form for time dependent coordinate 
systems:
\begin{equation} \label{flux_cons} 
 \sqrt{|\mathbf{g}|} \mathbf{K}  = \partial_t
\left(\sqrt{|\mathbf{g}|} \mathbf{d} \right) + \partial_{i}
\left(\sqrt{|\mathbf{g}|} \mathbf{{{e}}}^{i} \cdot (\mathbf{u} -
\dot{\mathbf{x}}) \mathbf{d} \right)
\end{equation}
For the full analytic derivation we refer to ~\citep{thompson+warsi_1985} again.
Defining the contravariant velocity components relative to the moving 
grid by $U^{i} = \mathbf{{{e}}}^{i} \cdot (\mathbf{u} - \dot{\mathbf{x}})
$ the above equations yield
\begin{equation}
 \sqrt{|\mathbf{g}|} \mathbf{K} = \partial_t
\left( \sqrt{|\mathbf{g}|} \mathbf{d} \right) + \partial_{i}
\left( \sqrt{|\mathbf{g}|} U^{i} \mathbf{d} \right) .
\end{equation}

\subsection{Set of RHD Equations in Strong Conservation Form} \label{full_set}

We exhibit the system of equations of radiation hydrodynamics in
somewhat simplified formulation. The following system has been basis for a number of implicit RHD 
computations (see e.g.~\citep{dorfi_1999}). All the astrophysical assumptions, implications  
and simplifications can be found in~\citep{mihalas+mihalas_1984}. In this paper we 
only want to emphasize the structural form of such a set of equations in a strong conservation 
form for comoving curvilinear coordinates. Note that for scalar equations the only effectively remaining 
geometric term inside the derivatives is the volume element $\sqrt{|\mathbf{g}|}$. The 
vectorial equations contain however also the time dependent base vectors.

\subsubsection{Continuity Equation}

The strong conservation form of the continuity equation 
\begin{equation*}
 \partial_t \rho + \mathrm{div} \, (\rho \mathbf{u} ) = 0 
\end{equation*}
is given for time depended coordinates by
\begin{equation*} 
 \partial_t
\left(\sqrt{|\mathbf{g}|} \rho \right) + \partial_{i}
\left(\sqrt{|\mathbf{g}|} \mathbf{{{e}}}^{i} \cdot (\mathbf{u} -
\dot{\mathbf{x}}) \rho \right) = 0.
\end{equation*}   

\subsubsection{Equation of Motion}

The equation of motion that we consider in radiation hydrodynamics contains 
the conservation of moment of plain fluid dynamics \eqref{hd_cons},
the radiative flux as a coupling term ($\mathbf{H}$, with the specific Rosseland opacity of the 
fluid $\kappa_R$), gravitational force ($\mathbf{G}$) and the artificial viscosity term ($\mathbf{Q}$ from \eqref{art_visc_corr}): 
\begin{equation*}
 \partial_t (\rho \mathbf{u}) + \mathrm{div} \, \big( \rho \mathbf{u}^{\mathrm{T}}
\mathbf{u} + \mathbf{P} + {\mathbf{Q}} \big) + \rho \mathbf{G} - \frac{4
\pi}{c} \kappa_R \rho 
\mathbf{H}  = 0.
\end{equation*}
We investigate the elements of the energy stress tensor a little 
closer before we give the consistent strong conservation form. 
We define an effective tensor of gaseous momentum $\mathbf{R}$
that accounts for 
the motion of the coordinates as
\begin{equation*}
 \mathbf{R} = r^{ij} \mathbf{{e}}_i \mathbf{{e}}_j
= \rho (\mathbf{u} - \dot{\mathbf{x}} ) \, \mathbf{u}^{\mathrm{T}}.
\end{equation*}
The isotropic gas pressure tensor $\mathbf{P}$ 
is defined by the scalar pressure and the metric tensor as 
$\mathbf{P} = p \mathbf{g}$. 
The viscous pressure tensor ${\mathbf{Q}}$ is to be modified in the way we suggested in 
\eqref{art_visc_corr}. Since in most applications of RHD with self-gravity involved, the gravitational 
force $\mathbf{G}$ is determined by solving the Poisson equation for the potential $\Phi$, we substitute 
$\mathbf{G} = - \nabla \Phi$. The equation of motion in strong conservation form is then written as
\begin{eqnarray*}
  \partial_t \left( \sqrt{|\mathbf{g}|} \rho \mathbf{u} \right) + \partial_i \left(
\sqrt{|\mathbf{g}|} \, \big( \mathbf{R} + \mathbf{P} + {\mathbf{Q}} \big) 
\cdot \mathbf{{{e}}}^i \right) \, + && \\ 
+ \, \rho \partial_i \left( \sqrt{|\mathbf{g}|} \Phi
\mathbf{{e}}^i \right) - \frac{4 \pi}{c} \kappa_R \sqrt{|\mathbf{g}|} \rho \mathbf{H}  = 0.
\end{eqnarray*}
The $k$th component of the strong conservation equation of motion is given
by the $k$th Cartesian  component of the unit vector, e.g. in spherical 
coordinates $\mathbf{{{e}}}^r = 
\cos\varphi \sin\vartheta \mathbf{e}^x + \sin\varphi \sin\vartheta \mathbf{e}^y 
+ \cos \vartheta \mathbf{e}^z$  
and its derivatives. The projection of each physical tensor on the contravariant coordinate lines can be
simplified by its contravariant components with respect to its covariant basis
without losing strong conservation form, i.e. $\mathbf{f}\cdot \mathbf{e}^i = 
f^{i,j} \mathbf{e}_j$. 
We prefer this form with contravariant 
components since it meets the native design of the stress tensor $\mathbf{R}$ and 
the pressure tensor $\mathbf{P}$. The tensor of artificial viscosity as given in \eqref{art_vis_corr_comp}
is brought to contravariant form by the summation
$$
Q^{ij} = Q_{lm}g^{li}g^{mj} = \dots = \left( g^{li}g^{mj}(\nabla_l u_m + \nabla_j u_m ) - g^{ij}
\frac{1}{3} \mathrm{div} \,  \mathbf{u} \right) 
$$
and then the equation of motion as 
\begin{equation} \label{EOM-comp}
\begin{aligned}
 \partial_t \left( \sqrt{|\mathbf{g}|} 
\rho u^i \mathbf{{e}}_i \right)
+ \partial_i \left(
\sqrt{|\mathbf{g}|} \left( r^{ij} + p^{ij} + Q^{ij} \right)
\mathbf{{{e}}}_j \right) + && \\
+ \, \rho \partial_i \left( \sqrt{|\mathbf{g}|} \Phi
\mathbf{{e}}^i
\right) - \frac{4 \pi}{c} \kappa_R \sqrt{|\mathbf{g}|} 
\rho H^i \mathbf{{e}}_i  = 0.
\end{aligned}
\end{equation}

\subsubsection{Equation of Internal Energy}

The energy equation
\begin{equation*}
 \partial_t (\rho \epsilon) + \mathrm{div} \, (\mathbf{u} \rho
\epsilon) + \mathbf{P} \cdot \nabla\mathbf{u}^{\mathrm{T}} - 4 \pi \kappa_P 
\rho (J-S) + \mathbf{Q} \cdot \nabla\mathbf{u}^{\mathrm{T}} = 0 
\end{equation*}
accounts for the
thermodynamics of the fluid, namely the energy balance including kinetic and pressure 
parts as well as inner energy. Latter is a thermodynamic quantity which is 
associated with the equation of state. The specific inner energy ($\epsilon$)
is in case of an ideal fluid its thermic energy. Another term comes from the energy exchange 
with the radiation field ($(J-S)$-term) containing the specific Planck opacity $\kappa_P$ 
and viscous energy dissipation, expressed by the contraction of 
the viscosity with the velocity gradient $\mathbf{Q} \cdot \nabla \mathbf{u}^{\mathrm{T}}$. 

Since we assume isotropic gas pressure $\mathbf{P}=p \mathbf{g}$ 
we can reformulate its contribution via the Ricci Lemma 
and obtain a very simple scalar expression. 
$$
\mathbf{P} \cdot \nabla\mathbf{u}^{\mathrm{T}} = g^{ij} p \nabla_i u_j = p \nabla_i u^i = p \, \mathrm{div} \, \mathbf{u}
$$
The viscous energy dissipation is given by the contraction
$$
\mathbf{Q} \cdot \nabla \mathbf{u}^{\mathrm{T}} =  Q^{ij} \nabla_i u_j =  Q^{ij} \left( \partial_i u_j - {\Gamma^k}_{ij} u_k \right)  =: \mathsf{E}_{\mathrm{diss}},
$$ 
and the strong conservative form of the energy equation is then given by 
\begin{equation*}
\begin{aligned}
\partial_t \left(\sqrt{|\mathbf{g}|} \rho \epsilon \right) + 
\partial_i \left( \sqrt{|\mathbf{g}|} \rho \epsilon \, 
\mathbf{{{e}}}^{i} \cdot (\mathbf{u} - \dot{\mathbf{x}}) \right) 
 + \sqrt{|\mathbf{g}|} p \, \mathrm{div} \, \mathbf{u} \, - && \\
\label{EIE_cons} 
- \, 4\pi \sqrt{|\mathbf{g}|} \kappa_P \rho (J-S) + \sqrt{|\mathbf{g}|}
\mathsf{E}_{\mathrm{diss}} = 0.
\end{aligned}
\end{equation*} 

\subsubsection{Equation of Radiation Energy}

We write a simplified frequency integrated radiation energy equation in the 
comoving frame as follows
$$
 \partial_t J + \mathrm{div} \,  
(\mathbf{u} J) + c \, \mathrm{div} \,  \mathbf{H} + 
\mathbf{K} \cdot \nabla\mathbf{u}^{\mathrm{T}} + c \chi_P (J-S) = 0.
$$
For the scalar energy input of radiative pressure into the material 
we define a new coupling variable 
$$
\mathbf{K} \cdot \nabla \mathbf{u}^{\mathrm{T}} = K^{ij} \nabla_i u_j =: \mathsf{P}_{\mathrm{coup}}
$$ 
and in strong conservation form the equation of radiation energy is given by
\begin{equation*}
\begin{aligned}
\partial_t\left(\sqrt{|\mathbf{g}|} J \right) + 
\partial_i \left( \sqrt{|\mathbf{g}|} \, \mathbf{{{e}}}^{i} \cdot \left( J 
(\mathbf{u}-\dot{\mathbf{x}}) +c \mathbf{H}\right) \right) + && \\
+ \, \sqrt{|\mathbf{g}|} 
\mathsf{P}_{\mathrm{coup}} + 
\sqrt{|\mathbf{g}|} c \chi_P (J-S) = 0.
\end{aligned}
\end{equation*}
\subsubsection{Radiative Flux Equation}

Another vectorial conservation law, the radiative flux equation, remains to be written:
$$
 \partial_t \mathbf{H} + \mathrm{div} \,  (\mathbf{u} \mathbf{H}) 
+ c \, \mathrm{div} \,  \mathbf{K} + 
\mathbf{H} \cdot \nabla\mathbf{u}^{\mathrm{T}} + c \chi_R
\mathbf{H} = 0.
$$
We define an effective radiative flux tensor $\mathbf{L}$ analogously to the effective tensor of gaseous momentum: 
$$
(\mathbf{u}-\dot{\mathbf{x}}) \mathbf{H} =: \mathbf{L} = l^{ij}  \mathbf{{e}}_i \mathbf{{e}}_j 
$$
and for the contribution of radiative momentum to the material 
$\mathbf{H} \cdot \nabla\mathbf{u}^{\mathrm{T}}$ we
define another coupling variable $\mathsf{F}$ with components 
$$
\nonumber \mathsf{F}_{\mathrm{coup}}^j = H^i \nabla_i u^j.
$$
The geometrically conservative form of the radiative flux equation in non-steady coordinates 
is then written as
\begin{equation*}
 \begin{aligned}
\partial_t \left( \sqrt{|\mathbf{g}|} \mathbf{H} \right) + \partial_i 
\left( \sqrt{|\mathbf{g}|} \, \mathbf{{{e}}}^{i} 
\cdot \left( \mathbf{L} + c \mathbf{K} \right) \right) + && \\
+\,  \sqrt{|\mathbf{g}|}
 \mathsf{F}_{\mathrm{coup}}
+ \sqrt{|\mathbf{g}|} \kappa_R \rho \mathbf{H} = 0 . 
\end{aligned}
\end{equation*}

\subsection{Acknowledgements}
We thank Franz Embacher and Helmuth Urbantke for valuable discussions on 
tensor analysis and differential geometry. 
The authors acknowledge the UniInfrastrukturprogramm des BMWF 
Forschungsprojekt Konsortium Hochleistungsrechnen, the 
Forschungsplattform Scientific Computing at LFU Innsbruck and the doctoral school - 
Computational Interdisciplinary Modelling FWF DK-plus (W1227).
This research employed resources of the Center for Computation and Technology at 
Louisiana State University, which is supported by funding from the Louisiana 
legislature's Information Technology Initiative.

\bibliography{visco,benger,graphics}
\renewcommand{\bibsection}{\section{References}}
\setlength{\bibhang}{1.24cm}
\setlength{\parindent}{3cm}
\setlength{\bibsep}{0cm}
\bibliographystyle{dcu}
\setcitestyle{authoryear,round,citesep={;},aysep={,},yysep={;}}
\gdef\harvardand{\&}
\end{document}